# CIIA: A New Algorithm for Community Detection

Zhang renquan, Wang yu, Wang xiaolin, Sun yuze, Tai jilei

College of Mathematics Science,

Dalian University of Technology, Dalian, 116000, China

Abstract: In this paper, through thinking on the modularity function that measures the standard of community division, a new algorithm for dividing communities is proposed, called the Connect Intensity Iteration algorithm, or CIIA for short. In this algorithm, a new indicator is proposed.This indicator is the difference between the actual number of edges between two nodes and the number of edges when the edges are randomly placed. It can reflect more information between the nodes. The larger the value of this index, the greater the possibility that the two nodes are divided into the same community, and vice versa. This paper also verifies the algorithm through numerical simulations and real cases, and the results show the feasibility of the algorithm.

## 1 Introduction

Community detection is one of the popular domains of complex network analysis and community research. It has been more than a hundred years since the problem was first proposed.Complex network is essentially a graph composed of nodes and edges. Traditional methods are mainly used to deal with Graph Segmentation and Clustering Problems[1, 2], mainly including hierarchical clustering, partition clustering, Graph Segmentation and spectral analysis[3] etc. These algorithms have increased attention to the problem of subgraph segmentation and community structural issues.

Complex networks can be divided into static networks and dynamic networks according to whether there is any subsequent change. Community discovery is usually to find a community that is as structured as possible under a given standard, that is, to optimize the objective function. The purpose of the algorithm is very clear: to find the best community in the fastest time. On static and dynamic networks, since there is no clear definition of the community structure, scholars use different standards and methods to divide communities, and thus many discovery algorithms are proposed.

The algorithms of static networks can be roughly divided into split algorithms, methods based on modularity, and methods based on statistical inference. The representative of the splitting algorithm is the GN algorithm[4, 5], which proposes the concept of edge betweenness to mark the influence of each edge on network connectivity. With the community structure measurement function proposed by Newman: modularity[6], many algorithms with optimized modularity as the objective function were born. Modularity maximization is an NP-hard problem, so we can only find an approximate solution. For example, greedy algorithm[7], Extremal optimization algorithm[8, 9], GA algorithm based on simulated annealing[10], flow coding algorithm[11, 12], etc. Methods based on statistical inference include observed data sets and model assumptions[13, 14].

Dynamic algorithms include faction filtering[15, 16], aggregation algorithm based on similarity[17, 18], and label propagation algorithm[19, 20]. This paper mainly studies static networks, without too much introduction to dynamic networks and dynamic algorithms.

Louvain algorithm is also a heuristic method based on modularity optimization[21], similar to the greedy algorithm[7], both are based on the maximization of local modularity. Compared with other community detection methods, it has greater advantages in terms of calculation speed, and can obtain a higher modularity, which has attracted wide attention. One of the reasons for the high efficiency of this algorithm is that by moving isolated nodes to new communities, the modularity gain $\Delta Q$ can be easily calculated. The algorithm uses the idea of graph reconstruction to achieve hierarchical clustering and does not need to manually set the number of communities, and only spend some time at the lowest layer of the algorithm. The algorithm proposed in this paper is also inspired by the idea of modularity[6]. Therefore, the Louvain algorithm is selected as the contrast, and the two indicators of time and modularity are compared to test the performance of the algorithm.

Modularity, as a common standard to measure the quality of the divided community structure.By calculating the difference between the actual weight of two nodes and the weight when randomly placed, the quality of the network community can be quantitatively measured. High modularity means that there are relatively dense edges between the vertices of the same community, and the edges between the vertices divided by different communities are relatively sparse.The more obvious the community structure of the network, the higher the modularity obtained.

However, after a lot of experiments, it is found that the modularity is the sum of global information. The calculation process relies on the final community belonging of two nodes. When only the actual weights of the two nodes i, j and the weights are randomly placed, the information is not enough to directly judge whether the two nodes are in the same community.It has been verified by experiments that given an edge, the size of the overlap between the two corresponding nodes and their neighboring points can make up for the above shortcomings, and can reflect whether the edge is an internal edge of the community.

## 2 Deduce

In the field of complex network community discovery, we often define $G$ to represent a network structure.The number of nodes and edges of network $G$ are respectively $n$ and $m$. Use $k_i$ to denote the degree of node $i$, and $k_{S_{i \to j}}$ to denote the sum of degrees of $S_{i \to j}$, where $S_{i \to j} = \{l | l \in \partial i \cup \{i\} \backslash j\}$, which represents the point set composed of point $i$ and its adjacent points (except j), which is called $i$ circle, and $j$ circle is the same. Both of these two degrees can be expressed in the form of adjacency matrix $A. k_i = \sum_{j \in G} A_{ij}, k_{S_{i \to j}} = \sum_{l \in S_{i \to j}, j \in G} A_{ij}$.

Taking point pair $i$, $j$ as an example, the relationship circle is influenced by each other. $CI_{ij}$, the difference between the actual connecting edge between two circles and the connecting edge when placed randomly, can be used as the basis for judging the close degree of connection between $i, j$, which is called connect intensity here. The overlapping parts of $i$ circle and $j$ circle are mainly affected by the common adjacent points and the mutual connecting edges between adjacent points.

The larger the $CI_{ij}$ is, the closer the connection is, and the more $i$ and j tend to be in the same community. After calculating the $CI_{\text{value}}$ for all edges, we only need to sort the $CI_{\text{value}}$ in reverse order, and then we can know the probability order that the pairs of points will eventually be in the same community. In order to verify the feasibility of the method, in this paper, like Louvain's method, calculates the modular gain $\Delta Q$ by moving isolated nodes to a new community, and merges regularly, but terminates if it is negative.

$$CI_{ij} = E_a^{ij} - E_p^{ij} = A_{S_{i \to j} S_{j \to i}} - \frac{k_{S_{i \to j}} k_{S_{j \to i}}}{2m}, \tag{2.1}$$

$E_a^{ij}$ means the actual connected edge, and $E_p^{ij}$ is the theoretical connected edge, $A_{S_{i \to j} S_{j \to i}}$ stands for the number of connected edges where $i$ circle and $j$ circle cross, $k_{S_{i \to j}}$ stands for the sum of degrees of all nodes in the $i$ circle, $m$ represents the total number of edges of the network. Now let us take an example as shown in Figure 1. Here, we have $S_{2 \to 4} = \{\{1,3,4,5,9,12\} \cup \{2\} \setminus 4\} = \{1,2,3,5,9,12\}$. The same, $S_{4 \to 2} = \{3,4,10\}$.

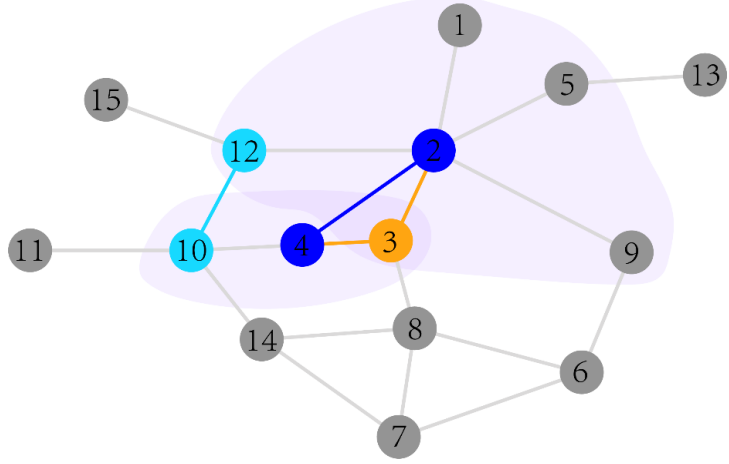

Figure 1: The relationship between the adjacent points of node 2 and 4 and its adjacent points. When discussing blue connected edges (2,4), we pay attention to the existence of such edges as (2,3), (3,4) and (10,12).

In the actual calculation process, due to the huge scale of the network and the overlapping part between $i$ circle and $j$ circle, in order to avoid repeated calculation, the Formula (2.1) is modified as follows: $E_{ra}^{ij}$, $E_{rp}^{ij}$ represent the actual connected edge and theoretical connected edge which have been calculated many times.

$$CI_{ij} = \left(E_a^{ij} - E_{ra}^{ij}\right) - \left(E_p^{ij} - E_{rp}^{ij}\right), \qquad (2.2)$$

$$E_a^{ij} = \sum_{u \in S_{i \to j}, v \in S_{j \to i}} A_{uv}, \qquad (2.3)$$

$$E_{ra}^{ij} = \frac{1}{2} \sum_{u,v \in (S_{i \to j} \cap S_{j \to i})} A_{uv}, \qquad (2.4)$$

$$E_p^{ij} = \frac{k_{S_{i \to j}} k_{S_{j \to i}}}{2m} = \frac{\sum_{u \in S_{i \to j}} k_u \sum_{v \in S_{j \to i}} k_v}{2m}, \qquad (2.5)$$

$$E_{rp}^{ij} = \frac{\sum_{u \in (S_{i \to j} \cap S_{j \to i})} k_u^2 + \frac{1}{2} \sum_{u,v \in (S_{i \to j} \cap S_{j \to i})} k_u k_v}{2m}, \qquad (2.6)$$

Still taking Figure.1 as the example, If we calculate the $CI_{24}$, which can be seen from the figure:

$$m = 20, E_a = 4, E_{ra} = 0;$$

$$k_{S_{2 \to 4}} = 17, k_{S_{4 \to 2}} = 10, E_p = \frac{17 \times 10}{2 \times 20} = 4.25,$$

$$S_{2\to 4} \cap S_{4\to 2} = \{3\}, k_3 = 3, E_{rp} = \frac{3^2}{2 \times 20} = 0.225$$

So, $CI_{24} = (4-0) - (4.25 - 0.225) = -0.025$. The $CI_{\text{value}}$ of each edge is given below, and the result is obtained by formula (2.2).

Table 1: $CI..$ of each edge

| Edge | $CI..$ | Edge | $CI..$ | Edge | $CI..$ | Edge | $CI..$ |
| --- | --- | --- | --- | --- | --- | --- | --- |
| (7, 8) | 2.75 | (13, 5) | 0.8 | (3, 4) | -0.325 | (9, 2) | -1.25 |
| (6, 7) | 2.15 | (12, 15) | 0.675 | (2, 5) | -0.35 | (10, 4) | -1.3 |
| (14, 7) | 1.65 | (10, 11) | 0.675 | (10, 12) | -0.75 | (12, 2) | -1.4 |
| (6, 8) | 1.625 | (1, 2) | 0.525 | (9, 6) | -1.0 | (14, 10) | -1.75 |
| (14, 8) | 0.975 | (2, 4) | -0.025 | (2, 3) | -1.025 | (3, 8) | -2.9 |

According to the $CI_{\text{value}}$ calculation provided above, we choose the order of points joining the community according to the size of the connection strength. Whether to join the community needs to be judged according to the gain of modularity. Comparing the results of the Louvain algorithm on this network, the two can get the same modularity. However, our algorithm can know more information than modularity, such as an indicator between the edges of two nodes. In addition, because the louvain algorithm randomly selects the initial node, it may cause the modularity of each calculation to be different, and our algorithm is stable in the calculation of modularity. Figure 2 shows the process of community detection, and the modularity value of the network is 0.398750.

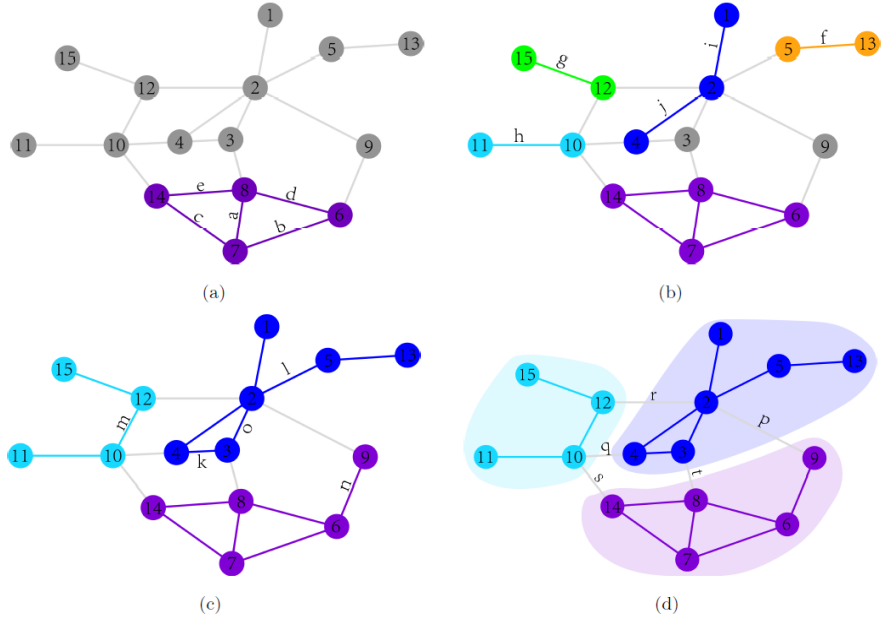

Figure 2: The process of community detection

Then, we carry out experiments on BA networks and ER networks, and the results show that our algorithm can run faster, The final modularities show that both our algorithm and Louvain method have their own advantages and disadvantages. Both BA and ER networks are generated by networkx package of Python. BA networks were generated by the function *barabasi-albert-graph(n, m, seed = None)*. The function will return a random graph according to the Barabási–Albert preferential attachment model. A graph of $n$ nodes is grown by attaching new nodes each with $m$ edges that are preferentially attached to existing nodes with high degree. Among them, we let $n$ ranges from 500 to 3000, and $m$ take the value 1. The reason why our algorithm performs well on these ER networks is that we strictly control the generation conditions of them. We use a function that can return the random partition graph with a partition of sizes. The function is random_partition_graph(*sizes, p_in, p_out, seed =None, directed = False*). Nodes in the same community are connected with probability $p\_in$ and nodes of different communities are connected with probability $p\_out$. We use this function to generate networks with an average degree of 6 and the probability of connecting edges within a group is more than one hundred times the probability of connecting edges between groups. The number of nodes ranges from 500 to 3000, and a network with 10 community structures is generated each time.

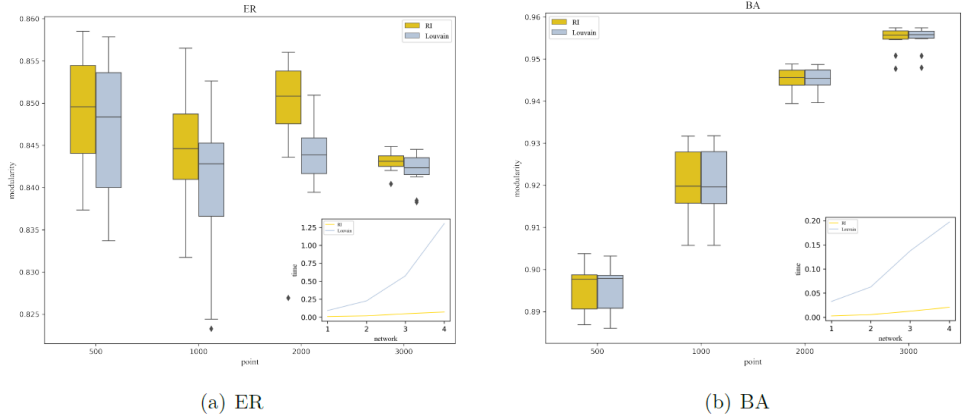

(a) ER  (b) BA

Figure 3: Numerical simulation. According to different sizes of BA and ER networks, CI algorithm and Louvain algorithm are used to compare the time and modularity.

## 3 Connect Intensity Iteration Algorithm Combined

Combined with the low modularity of CI algorithm in SF network, after exploring the reasons, it is found that in the circle selection of nodes, the final division results of some nodes adjacent points are not in the same community as this node, so such nodes should not be counted in the circle. Is there a way to find out only the nodes that tend to be divided into the same community with $i$ among the adjacent points of $i$?

The ratio of $CI..$ calculated in the previous round can be used as the weight in the next round, and continuous training can be achieved. The iterative process is the process of self-learning and self-selection. Assuming that $CI_{il}$ of node $i$ and node $j$ are negative in the first round of calculation, we think that they tend to be disconnected. Let the weight between them be 0 in the next round. If the $CI..$ of node $u$, $v$ are positive in the first round of calculation, the weight of $u$ to $v$ is the ratio of their $CI..$ to all $CI..$ of $u$, and the formula is as follows:

$$CI'_{ui} = \begin{cases} CI_{ui}, & CI_{ui} > 0 \\ 0 & otherwise \end{cases}, \quad w_{ui} = \frac{CI_{uv}}{\sum_{v \in \partial u} CI_{uv}}, \tag{3.1}$$

It should be noted that although $CI_{ui} = CI_{iu}$, $w_{ui} \neq w_{iu}$ is due to the difference of $\sum_{v \in \partial u} CI_{uv}$, Then, for formula (2.3) to formula (2.6), corresponding modifications can be made:

$$E_a^{ij} = \sum_{u \in S_{i \to j}, v \in S_{j \to i}} w_{ui} \, w_{vj} A_{uv}, \tag{3.2}$$

$$E_{ra}^{ij} = \frac{1}{2} \sum_{u,v \in (S_{i \to j} \cap S_{j \to i})} \frac{w_{ui} w_{vj} + w_{vi} w_{uj}}{2} A_{uv}, \qquad (3.3)$$

$$E_{p}^{ij} = \frac{k_{S_{i \to j}} k_{S_{j \to i}}}{2m} = \frac{\sum_{u \in S_{i \to j}} w_{ui} k_u \sum_{v \in S_{j \to i}} w_{vj} k_v}{2m}, \qquad (3.4)$$

$$E_{rp}^{ij} = \frac{\sum_{u \in (S_{i \to j} \cap S_{j \to i})} \frac{w_{ui} + w_{uj}}{2} k_u^2 + \frac{1}{2} \sum_{u,v \in (S_{i \to j} \cap S_{j \to i})} \frac{w_{ui} w_{vj} + w_{vi} w_{uj}}{2} k_u k_v}{2m}, \qquad (3.5)$$

Now we give another example to illustrate the improved algorithm, as shown in Figure 4(e). We want to calculate the $CI..$ between node 7 and node 8. From the previous section, we know that before the algorithm is improved, we have $CI_{78} = -1.21875$. But after a few iterations, it can be calculated that $CI_{78}^{(1)} = 0.3901$, $CI_{78}^{(2)} = 0.29974$, $CI_{78}^{(3)} = 0.30419$. The $CI..$ of each edge calculated by formula (2.2) is given below，and the modularity value of this network is 0.21875 before the algorithm is improved. But after iterating through formulas (3.1) to (3.5), the modularity can finally reach 0.283203125. Table 2 shows the $CI..$ of edges after each iteration, and Figure 5 shows the division process after the final iteration.

According to the ranking results of each round, the modular degree after iteration is 0.283203125, which is consistent with the result of Louvain's division. This division method divides the 16 edges in the network into 10 community inner edges and 6 community outer edges. It can be clearly seen in the table below that with the iteration process, the $CI..$ of community inner edges gradually widens the gap with the $CI..$ of even edges between communities, and finally can be distinguished. Therefore, the original idea of this paper has been achieved. When the community is merged into a connected edge with negative modularity gain, the subsequent division can be stopped.

Table 2: The $CI..$ of each edge after teration

| Original | | 1st Iteration | | 2nd Iteration | | 3rd Iteration | |
|---|---|---|---|---|---|---|---|
| edge | $CI..^0$ | edge | $CI..^1$ | edge | $CI..^2$ | edge | $CI..^3$ |
| (5, 6) | 1.625 | (5, 6) | 1.4375 | (5, 6) | 1.3478 | (5, 6) | 1.4985 |
| (0, 8) | 1.53125 | (5, 9) | 1.3953 | (6, 9) | 1.2693 | (6, 9) | 1.4637 |
| (6, 9) | 0.78125 | (6, 9) | 1.375 | (5, 9) | 1.2098 | (5, 9) | 1.4053 |
| (0, 1) | 0.65625 | (3, 7) | 1.3149 | (3, 4) | 1.0987 | (3, 4) | 1.1745 |
| (1, 8) | 0.65625 | (3, 4) | 1.1264 | (4, 7) | 1.0104 | (3, 7) | 1.114 |
| (4, 7) | 0.46875 | (0, 8) | 1.0619 | (0, 8) | 1.0026 | (4, 7) | 1.1129 |
| (1, 2) | 0.40625 | (1, 2) | 0.781 | (3, 7) | 0.99 | (0, 8) | 1.0255 |
| (0, 7) | 0.25 | (1, 8) | 0.6931 | (1, 2) | 0.7751 | (1, 2) | 0.7756 |
| (3, 4) | -0.25 | (4, 7) | 0.5962 | (0, 1) | 0.6307 | (0, 1) | 0.6716 |
| (5, 9) | -0.625 | (3, 5) | 0.5921 | (1, 8) | 0.6045 | (1, 8) | 0.5899 |
| (3, 7) | -0.65625 | (0, 1) | 0.5913 | (0, 7) | 0.443 | (0, 7) | 0.3627 |
| (7, 8) | -1.21875 | (0, 7) | 0.3955 | (7, 8) | 0.2997 | (7, 8) | 0.3042 |
| (3, 5) | -1.8125 | (7, 8) | 0.3901 | (8, 9) | -0.1438 | (3, 5) | -0.1228 |
| (1, 9) | -2.0 | (8, 9) | 0.2495 | (1, 4) | -0.1525 | (8, 9) | -0.1746 |
| (8, 9) | -2.21875 | (1, 9) | 0.1365 | (1, 9) | -0.2203 | (1, 9) | -0.3161 |
| (1, 4) | -2.3125 | (1, 4) | -0.0323 | (3, 5) | -0.2508 | (1, 4) | -0.3181 |

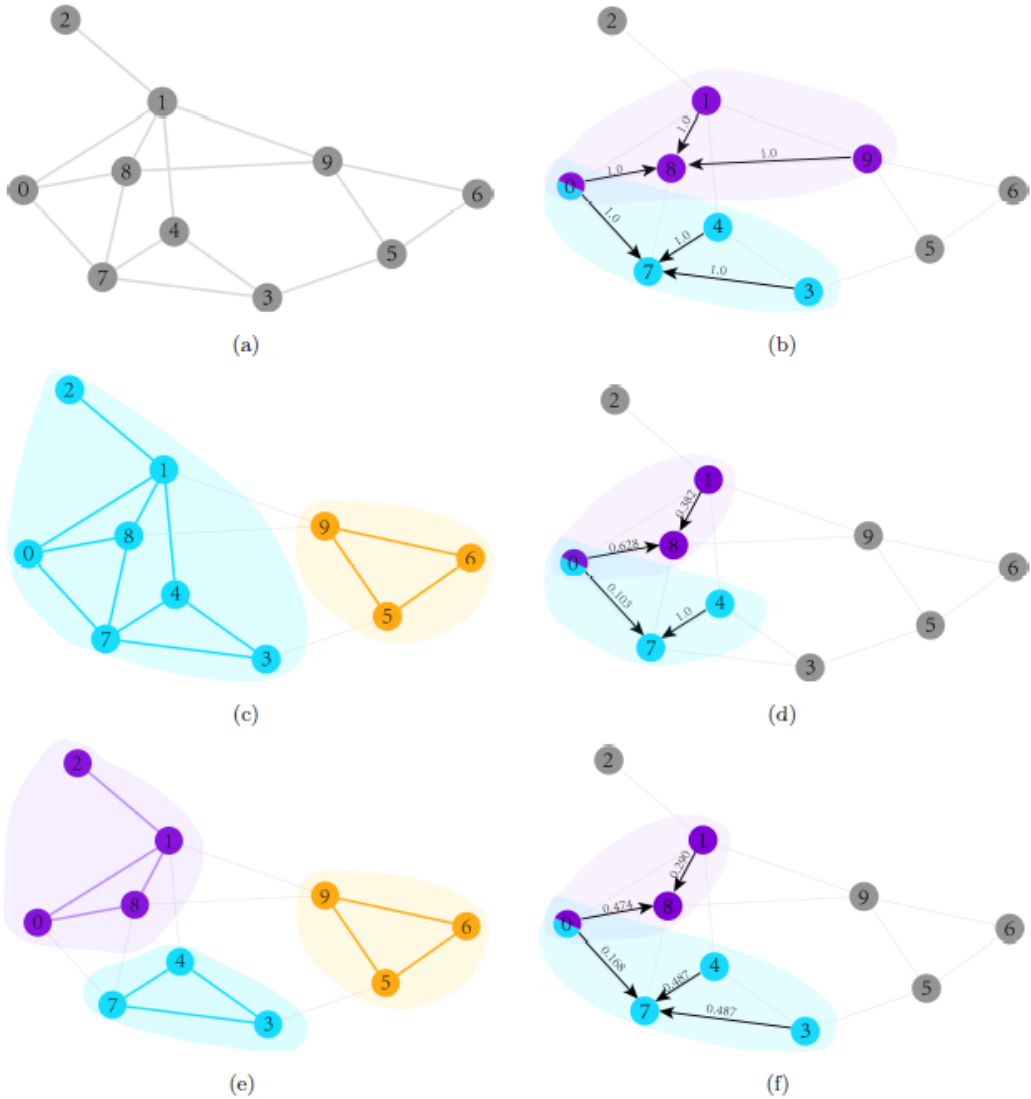

Figure 4: The process of community detection for Example 2. (a) represents the initial network, (c) is the result before iteration and (e) the division result after iteration.(b)(e)(f) represents the change of $S_{7 \to 8}$ and $S_{8 \to 7}$

### 3.1 Connect Intensity Iteration Algorithm steps

Here, we give the steps of CIIA algorithm. CI algorithm can skip steps 2 to 4.
(1) calculating the $CI..$ of each edge and taking each point as a community alone;
(2) According to the $CI..$, the weight w of each point and adjacent points is calculated, where the $w = 0$ of $CI.. < 0$ or $CI.. = 0$, and the rest is the proportion;
(3) according to the weighted node degree in step 2, calculate the connection strength again;
(4) Repeat steps 2 and 3 until the final $CI..$ sequence and sign have no fluctuation compared with the previous result;
(5) Merging the communities at the two ends of the corresponding side according to the CI.. from large to small (if $CI..$ is equal, it is random). If the modularity gain is positive, accept this merger to get a new partition. The next merger will be carried out on this new partition. When negative modularity gain occurs, the subsequent partition can be stopped.

# 4 Experiment

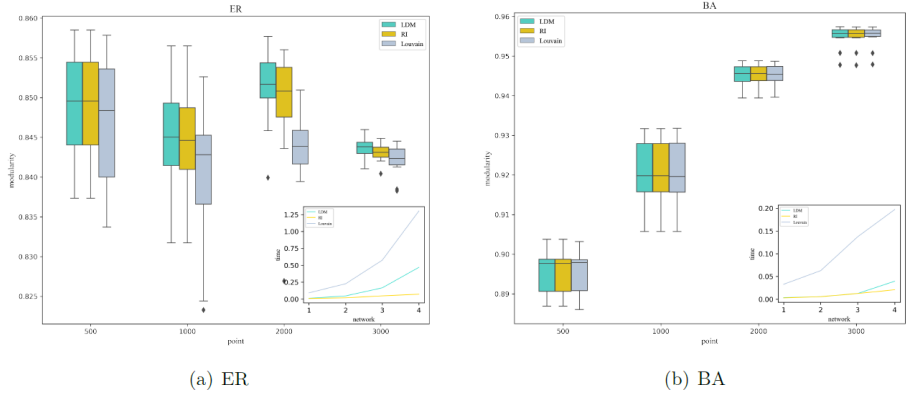

(a) ER

(b) BA

Figure 5: Still the data in Figure 3. According to different sizes of BA and ER networks, CIIA algorithm, CI algorithm and Louvain algorithm are used to compare the time and modularity index.

When we put the comparison results of the three algorithms together, we can find that compared with CI algorithm, CIIA algorithm has improved the modularity index of different networks after sacrificing a certain amount of time for iterative operation.Finally, based on the real network data published on open source data set websites such as Stanford University Data Network, the algorithm is tested, and the results are shown in Table 3:

Table 3: Real Networks

| info | Florentine[22] | Les Misèrables[23] | web-edu |
| --- | --- | --- | --- |
| nodes/edges | 15/20 | 77/254 | 3k/6.5k |
| CI | 0.3987/0.7ms | 0.5485/6.8ms | 0.9479/0.2s |
| Louvain | 0.3979/1.6ms | 0.5527/9.2ms | 0.9505/1.4s |
| CIIA | 0.3987/1.3ms | 0.5539/10.4ms | 0.9505/2s |

# 5 Conclusion

Through a large number of experiments, the algorithm in this paper has achieved good results on many networks. Compared with Louvain algorithm, it has three advantages:
1.The randomness of the initial selection of Louvain algorithm, the results of each division are different. But the results of any iteration run of CIIA algorithm remain unchanged.
2.CIIA algorithm can get more global information, and sometimes it can get higher modularity than Louvian algorithm.
3.Since the calculation time of the CIIA algorithm is greatly affected by the number of edges and average degree in the network, and less affected by the number of nodes, the speed of the sparse network is much faster than that of the Louvain algorithm.

The original idea of the method proposed in this paper is to sort the edges by the new index value, so that all the internal edges of the community under the optimized score are arranged before the edges between the communities, so that the process of calculating the modularity gain can be directly omitted. Through a lot of experiments, we found that this can be achieved on a small network. In a medium-sized network with more than 1000 nodes, in the edges ranking obtained by the new index value, about the top 20% and bottom 10% of the edges are the final internal and external edges. There are still some edges whose ordering fluctuates. Nevertheless, in the calculation of the final modularity, it has reached the 96%-105% accuracy of the Louvain algorithm. That is, in some cases, for example, the ER network can get a better division result than the classic algorithm, which shows that the research idea of the algorithm is correct.

Because when the edges with the strongest Relation Intensity are merged continuously, the communities corresponding to the two ends of the edges are merged. That is, the node sets are merged. In a sense, CIIA algorithm can not move a single node from one community to another to calculate and improve modularity. Therefore, this algorithm still has room for improvement.